\documentclass[journal=jacsat,manuscript=article]{achemso}

\usepackage[version=3]{mhchem} 
\usepackage{booktabs}
\usepackage[table]{xcolor}
\usepackage{amssymb}

\author{Gabriele Amante}
\affiliation{Department of Chemical, Biological, Pharmaceutical, and Environmental Sciences, University of Messina, Viale F. Stagno d’Alcontres 31, 98166 Messina, Italy}
\alsoaffiliation{Institute for Chemical-Physical Processes, National Research Council (CNR-IPCF), V.le F. Stagno d'Alcontres 37, 98158 Messina, Italy}

\author{Fortunata Panzera}
\affiliation{Department of Chemical, Biological, Pharmaceutical, and Environmental Sciences, University of Messina, Viale F. Stagno d’Alcontres 31, 98166 Messina, Italy}
\alsoaffiliation{Laboratoire de physique de L’École normale supérieure de Paris, CNRS, École Normale Supérieure Université Paris Sciences et Lettres, Sorbonne Université, Université de Paris, Paris 75005, France}
\alsoaffiliation{Institute for Chemical-Physical Processes, National Research Council (CNR-IPCF), V.le F. Stagno d'Alcontres 37, 98158 Messina, Italy}

\author{Gabriele Centi}
\affiliation{Department of Chemical, Biological, Pharmaceutical, and Environmental Sciences, University of Messina, Viale F. Stagno d’Alcontres 31, 98166 Messina, Italy}

\author{Jing Xie}
\affiliation{Ministry of Education Key Laboratory of Cluster Science, Beijing Key Laboratory of Photoelectronic/Electrophotonic Conversion Materials, School of Chemistry and Chemical Engineering, Beijing Institute of Technology, Beijing 100081 (P. R. China)
}

\author{Ali Hassanali}
\affiliation{International Centre for Theoretical Physics (ICTP), Strada Costiera 11, 34151 Trieste, Italy}

\author{A. Marco Saitta}
\affiliation{Laboratoire de physique de L’École normale supérieure de Paris, CNRS, École Normale Supérieure Université Paris Sciences et Lettres, Sorbonne Université, Université de Paris, Paris 75005, France}

\author{Giuseppe Cassone}
\affiliation{Institute for Chemical-Physical Processes, National Research Council (CNR-IPCF), V.le F. Stagno d'Alcontres 37, 98158 Messina, Italy}
\email{giuseppe.cassone@cnr.it}

\title[]
  {Interfacial Electric Fields in Water Nanodroplets are Weakly Dependent on Curvature and pH} 
  
\abbreviations{IR,NMR,UV}

\begin{document}

\begin{abstract}
The origin of enhanced reactivity in aqueous microdroplets remains debated, with interfacial electric fields (IEFs) often invoked as catalytic drivers. Here, we provide a quantum-mechanical, spatially resolved characterization of the electric field at air-water interfaces by combining deep-learning molecular dynamics with \emph{ab initio} re-sampling. Across planar interfaces and nanodroplets of varying curvature and charge state, we find an outward-oriented field of $\sim 1.0$--$1.2$ V/{\AA} along the intrinsic surface normal. Crucially, its magnitude scales linearly with the average number of hydrogen bonds per interfacial molecule, directly tying the field to the local hydrogen-bond network. Despite its large magnitude and contrary to common expectations, we find that curvature and pH exert only a minor influence on the IEF, becoming negligible at experimentally relevant droplet sizes and pH. Consequently, the reactivity differences observed in $\mu$m-sized droplets cannot be ascribed to variations in the IEF, which changes by a factor of only $\sim10^{-5}$ between $3$ and $40\mu$m-sized droplets. Moreover, the IEF is localized inside the interfacial region and rapidly vanishes within a few {\AA}. This strong spatial confinement renders the IEF strongly tied to the local electronic structure, identifying it as a local property of the air-water boundary rather than an independent physical driver of ``on-water'' catalysis.
\end{abstract}

\section{Introduction}

Over the past decade, the chemistry of water microdroplets has attracted significant attention, as reactions within them often proceed along pathways and rates markedly different from bulk aqueous solutions \cite{Narayan_Angewandte2005,RuizLopez2020MolecularReactionsAqueous,Prasoon2023,Kusaka2021}. Experiments have shown that processes such as urea synthesis from \ce{CO2} and \ce{NH3},\cite{Mohajer2025UreaDroplets} sulfate oxidation,\cite{liu2024chem} and ammonia formation from \ce{N2}\cite{Song2023AmmoniaMicrodroplets} can occur under ambient conditions with unexpectedly high efficiencies. These findings underpin the concept of ``on-water'' catalysis, whereby the air-water interface provides a chemically distinct environment that promotes transformations unfavorable in the bulk.\cite{Holden2025MechanismsChargedMicrodroplets}

Understanding the molecular origins of such behavior remains an open challenge\cite{LaCour_JACS25}. The droplet interface combines several features absent in homogeneous aqueous systems, including partial solvation, strong density gradients, and enhanced molecular anisotropy. Among the mechanisms proposed to rationalize microdroplet reactivity, particular attention has been devoted to the role of electric fields at aqueous interfaces\cite{ChanCHEMSOCREV2026}. Several studies have suggested that the air-water boundary may sustain electric fields on the order of $\sim0.2$ V/{\AA},\cite{LaCour_JACS25,zare_jpcl2020,Headgordon_NatCommun2022,CostaLopezJACS2023,liu2024chem} a magnitude that could potentially influence chemical reactions by stabilizing charge-separated states or lowering activation barriers. This hypothesis conceptually draws support from numerous computational \cite{Shaik2016,shaik_accchemres_2025,cassone_chemsci2017,cassone_jpcc2019,amante_jcp25,che_acscatal_2018,Laporte2015,MillerPNAS2014,LaporteJPCC2020} and experimental \cite{Aragones2016,Stuve,Huang} studies demonstrating that externally applied and \emph{adequately oriented static} electric fields can profoundly modify chemical reactivity in both homogeneous\cite{cassone_chemsci2017,Cassone_JPCL2020} and heterogeneous\cite{Creazzo_PCCP2020} aqueous environments. By analogy, spontaneous electrostatic fields generated at the droplet interface have been proposed to act as catalytic driving forces for microdroplet chemistry.\cite{LaCour_JACS25,FranciscoPNAS25,LaCount2025ElectricFieldsInterfaces,Headgordon_NatCommun2022,Lee_pnas_zare2019,Song2023AmmoniaMicrodroplets,liu2024chem}

Droplet curvature has also very recently been invoked as a potential contributor to interfacial electrostatics. Because decreasing droplet size alters the organization of surface water molecules and enhances the relative importance of the interface, several works have suggested that nano-to-micro scale curvature may amplify interfacial electric fields (IEFs) and modify the local electrostatic environment experienced by molecules at the surface increasing the catalytic properties of air-water interfaces.\cite{XiaPNAS25,ZareJACS25} In addition, interfacial water exhibits complex acid-base and charge-transfer properties that can lead to non-trivial pH and ion distributions across the air-water boundary.\cite{Mishra2012BronstedBasicity,Zhang_ACSPhys2024,Xing_angew2022,Wang2025ChargeTransfer} The behavior of ions at aqueous interfaces is itself subtle: specific-ion effects arising from differences in hydration, size, and polarizability can significantly perturb interfacial structure and electrostatics.\cite{Jungwirth2006,McCaffreyPNAS2017,SekiJACS2023} At the same time, the possible involvement of reactive oxygen species, ozone contamination, or droplet charging processes complicates the interpretation of microdroplet experiments.\cite{Mishra2012BronstedBasicity,Holden2025MechanismsChargedMicrodroplets,LaCour_JACS25}

However, the catalytic role of IEFs as well as their magnitude remain subject of intense debate. Several recent studies have questioned whether such fields are sufficiently large or persistent to account for the reactivity observed in microdroplets. Mishra and co-workers, for example, reported no detectable role of IEFs in the formation of H$_2$O$_2$.\cite{Musskopf_jpcl21,Gallo_ChemSci22,Eatoo24,Eatoo25} Gong \textit{et al.}\ found that, for a prototypical Diels-Alder reaction, catalytic effects arising from interfacial fields were negligible compared with confinement and evaporation-induced enrichment.\cite{Gong_JACS24} Colussi has similarly argued that internal fields in aerosolized droplets are considerably weaker than previously proposed\cite{Colussi2025PhysicalChemistryMicrodroplets}. 
Consistently, accurate quantum-chemical calculations indicate that electric fields of $\sim0.2$ V/{\AA} only modestly perturb charge transfer in hydrogen-bonds\cite{AmadeoJPCA25} and redox energetics\cite{MartinsCosta2025EffectElectricFields_ANIE}. 
Earlier classical electrostatic simulations also concluded that enhanced interfacial reactivity cannot be explained solely by electric field magnitude.\cite{MartinsCosta2023ElectrostaticsAirWater}
Recent vibrational sum-frequency generation (SFG) measurements by Bonn and co-workers at the planar air-water interface revealed no spectroscopic signatures consistent with exceptionally strong electric fields.\cite{Shirley2025Reevaluating} Besides, classical molecular dynamics simulations by Hassanali and co-workers demonstrated that the electric field experienced by the phenol O-H group at the interface is essentially governed by its immediate hydration environment.\cite{Solana_JCP25} 

Despite these important advances, obtaining reliable estimates of IEFs remains challenging. Many theoretical approaches rely on classical electrostatic models with fixed charges, thereby neglecting polarization effects, while spectroscopic studies often infer field strengths only indirectly via Stark frequency-field correlations.\cite{Boxer} 
Here, we address these limitations by combining neural-network-based deep-learning molecular dynamics with \emph{ab initio} re-sampling, enabling an accurate and spatially resolved quantum-mechanical characterization of the electric field at both flat and curved air-water interfaces in nanodroplets of varying size, under neutral and charged conditions. Contrary to common expectations, we find that curvature and pH exert only a minor influence on the IEF, indicating that the enhanced reactivity of these environments must originate from alternative molecular mechanisms.

\section{Methods}

Deep Potential Molecular Dynamics (DPMD) simulations of air-water systems were carried out with the LAMMPS software (v. 29Aug2024) \cite{LAMMPS}. The reactive DP trained on the hybrid meta-GGA M06-2X \cite{M062X} exchange and correlation functional reported in Ref. \cite{Zhang_Langmuir25} was employed. Periodic simulations were carried out to mimic the behavior of the flat air-water interface, whereas non-periodic simulations of isolated nanodroplets were performed to probe the impact of the interfacial curvature on the surface electric field. As for the flat interface, samples composed of 256 and 512 H$_2$O molecules were considered with box edges of $a=b=19.641,c=45.000$ {\AA} and $a=b=24.830,c=60.000$ {\AA}, respectively. Seven isolated nanodroplets composed of 256, 384, 512, 640, 768, 896, and 1024 H$_2$O molecules were simulated in cubic boxes with edges of 44, 48, 60, 63, 65, 67, and 68 {\AA}, respectively, whereas the respective radii $R$ and the curvature $\kappa$ are listed in Table S1 of the Supporting Information (SI). Also charged nanodroplets composed of 507 H$_2$O and 5 OH$^-$ ($q=-5e^-$), 511 H$_2$O and 1 OH$^-$ ($q=-e^-$), 511 H$_2$O and 1 H$_3$O$^+$ ($q=+e^-$), and 507 H$_2$O and 5 H$_3$O$^+$ ($q=+5e^-$) were simulated, in addition to a globally neutral nanodroplet composed of 502 H$_2$O, 5 OH$^-$, and 5 H$_3$O$^+$ which, however, underwent charge recombination. Each DPMD simulation was carried out in the NVT canonical ensemble for at least 1 ns. The temperature was kept fixed at $300$~K via the canonical-sampling-velocity-rescaling (CSVR) thermostat~\cite{CSVR} with a damping factor of 0.04 ps. 

500 randomly chosen configurations were extracted from all DPMD trajectories and resampled via \emph{ab initio} calculations at the PBE+D3 \cite{PBE,Grimme1,Grimme2} Density Functional Theory (DFT) level to obtain the electric field $\mathbf{E}(\mathbf r)$ directly from the electron density $\rho_{e}(\mathbf r)$ via Poisson's equation:
\[
\nabla^2 \phi(\mathbf r)=-4\pi{\rho_{e}(\mathbf r)}
, \qquad
\mathbf E(\mathbf r)=-\nabla\phi(\mathbf r),
\]
where $\phi(\mathbf r)$ is the electrostatic potential. Although GGA exchange–correlation functionals are formally affected by delocalization error \cite{delocalization_error}, its impact on the present analysis is quantitatively negligible in the regime of interest. In particular, the interfacial electric fields (IEFs) reported here are intrinsically large (on the order of $\sim1$ V/{\AA}), such that any residual error associated with density delocalization is substantially smaller than the variations we discuss. This assessment is directly supported by benchmark calculations performed at the SCAN functional \cite{SCAN} and r2SCAN functional \cite{r2SCANerratum} levels on both air-water slabs and nanodroplets, which yield essentially indistinguishable IEF distributions (see Fig. S8 of the SI). Taken together, these results demonstrate that the magnitude and trends of the interfacial fields are robust with respect to the underlying density functional approximation, and that delocalization errors do not affect the physical conclusions drawn in this work. Due to their non-periodic nature and the large size of the vacuum regions, SCAN and r2SCAN calculations of nanodroplets containing more than $128$ H$_2$O species were not computationally feasible on computing nodes with $384$ Gb RAM. Electronic-structure calculations were carried out by employing the CP2K software (v. 2024.3) \cite{CP2K}. Wavefunctions of the atomic species have been expanded in TZVP basis sets with Goedecker-Teter-Hutter (GTH) pseudopotentials using the GPW method~\cite{GTH3}. Whereas for the PBE+D3 calculations a planewave cutoff of 400 Ry was imposed, for the SCAN and r2SCAN calculations a cutoff of 1200 Ry was adopted.

After computing the electric field vector on dense spatial grids of meshes in the range $\sim0.10-0.15$ {\AA}, subsequent analyses were performed using in-house C++ codes. For each configuration extracted from the DPMD simulations, the instantaneous Willard-Chandler (WC) interface was determined from the coarse-grained molecular density following the procedure introduced in Ref.~\cite{Willard2010InstantaneousInterface}. 
The intrinsic interface was then identified as the isodensity surface corresponding to a half of the bulk density. Using this surface, the simulation space was partitioned into distinct spatial regions according to the distance from the WC interface. As usual, the depth thresholds for discriminating between the bulk and the air-water interfacial regions were identified via the density profiles reported in the SI (Fig. S1). For each grid point of the electric field, the corresponding normal vector to the intrinsic surface, $\boldsymbol{n}_{\mathrm{int}}$, was computed. The electric field vector $\boldsymbol{E} = (E_x,E_y,E_z)$ was then projected onto the local WC reference frame and the normal component of the electric field was obtained as $E_n = \boldsymbol{E} \cdot \boldsymbol{n}_{\mathrm{int}}$.
Finally, the projected field components were accumulated over all grid points belonging to the same spatial region, allowing us to compute the average electric field along the interface normal in the bulk, interfacial, and vapors regions.

\section{Results and discussion}

\subsection{Structural and Electronic Properties of Air-Water Interfaces}

\begin{figure}[h]
\centering
\includegraphics[width=0.9\linewidth]{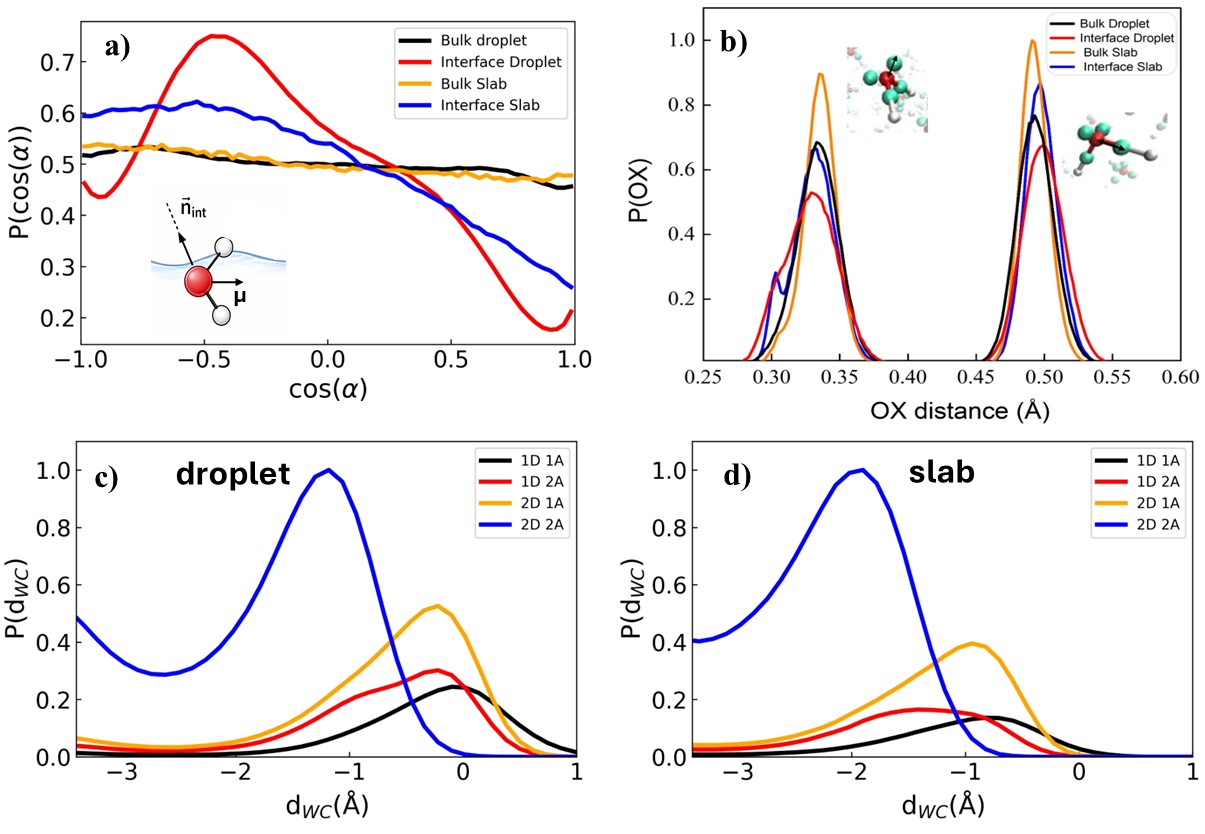}
\caption{\scriptsize Orientation distributions of molecular dipole moments P($\cos{\alpha}$) (a) with respect to the Willard-Chandler surface normal determined in the bulk and at the interfaces of a nanodroplet and a flat slab system composed of 512 H$_2$O molecules (see legend). In the respective inset, involved vectors are illustrated. Panel (b) reports the distribution of the O-X distance, where X denotes the position of the Maximally Localized Wannier Functions centers (visualized in cyan in the insets), for bulk and interfacial molecules in the droplet and slab geometries (see legends). Panels (c) and (d) show the spatial distribution of selected hydrogen-bond donor-acceptor (D-A) families as a function of the distance from the instantaneous Willard-Chandler interface $d_{\mathrm{WC}}$. Curves correspond to different hydrogen-bond topologies: 1D-1A (black solid), 1D-2A (red dashed), 2D-1A (orange dotted), and 2D-2A (blue dash-dotted). See Fig. S7 of the SI for the wider spectrum of hydrogen-bonded molecular families.}
\label{Figure1}
\end{figure}

The microscopic behavior of H$_2$O molecules at the air-water boundary differs markedly from that in the bulk. While molecular orientations are isotropic in the bulk, pronounced anisotropies arise at the liquid-vapor interface \cite{LaageJCP24}. Additional changes may occur with increasing curvature $\kappa = R^{-1}$, from planar slabs to nanodroplets of decreasing size. Fig.~\ref{Figure1}a reports the orientational distributions P($\cos{\alpha}$) of water dipole moments relative to the instantaneous Willard-Chandler (WC) surface normal, for both interfacial and bulk molecules. Whereas bulk regions exhibit nearly random distributions, distinct orientational populations emerge at the interface.

In the negative $\cos{\alpha}$ region, interfacial dipoles preferentially point toward the liquid phase, reflecting asymmetric cohesive interactions. The distribution maximum displayed in Fig.~\ref{Figure1}a around $\cos{\alpha} \sim -0.45$ indicates configurations that tend to maximize hydrogen bonding. 
In the positive $\cos{\alpha}$ region, corresponding to dipoles pointing toward the vapor phase, the slab distribution decays monotonically, whereas the droplet shows a shoulder around $\cos{\alpha} \sim 0.25$. This feature arises from configurations in which one OH bond remains oriented toward the liquid, stabilizing partially outward-pointing dipoles. In fact, by inspecting the orientations of the OH covalent bonds identified by the angle $\beta$ formed with the WC surface normal reported in Fig. S3b of the SI, a similar shoulder appears at $\cos{\beta} \sim -0.25$. Thus, a partial orientation of the dipole moment toward the vapor phase ($\alpha \sim 75^{\circ}$) is stabilized by one of the OH bonds of the same molecule pointing toward the interior of the liquid phase ($\beta \sim 105^{\circ}$).

Overall, the air-water interface of curved nanodroplets exhibits a more heterogeneous population of molecular orientations than flat interfaces, reflecting the larger number of H-bonds formed by water molecules in the planar slab, an aspect that will be discussed later. Within the same lines, de la Puente and Laage \cite{LaageJCP24} have also reported a larger number of free OH groups in a small nanodroplet of 128 H$_2$O with respect to a flat slab. Nevertheless, the fractions of outward- and inward-oriented molecules remain nearly identical in the droplet and flat slab (Figs. S3c,d). The excess of dipoles pointing toward the bulk, leading to an average $\alpha\sim{96^{\circ}}$ in the droplet ($\alpha\sim{94^{\circ}}$ in the slab), implies a net electric field directed outward from the liquid toward the vapor. 

The liquid-vapor interface is a strongly anisotropic environment in which the hydrogen-bond network is partially disrupted, progressively destabilizing tetrahedral coordination and leading to different polarization states with respect to the bulk. This is reflected in the O-X distributions (Fig.~\ref{Figure1}b), where X denotes the positions of the Maximally Localized Wannier Functions (MLWF)\cite{MLWF2} centers and O is the oxygen atom of a given H$_2$O. Each water molecule exhibits four MLWF centers: two associated with OH covalent bonds [$d(OX)\sim 0.5$~\AA] and two corresponding to lone pairs [$d(OX)\sim 0.35$~\AA]. Interfacial molecules display broader and slightly shifted distributions relative to the bulk, indicating increased electronic asymmetry. A distinct population at short O-X distances ($\sim 0.30$~\AA) -- visible in the curved nanodroplet interface (dashed red), but especially evident in the flat slab surface (dash-dotted blue) -- signals molecules with reduced hydrogen-bond acceptor capability.

This structural heterogeneity is further elucidated in Figs.~\ref{Figure1}c,d, which report the distribution of hydrogen-bond donor-acceptor (D-A) families as a function of the distance from the WC interface. While the bulk is overwhelmingly dominated by tetrahedral 2D-2A configurations, the interfacial region exhibits a pronounced enrichment of under-coordinated species such as 2D-1A and 1D-2A, reflecting asymmetric and partially disrupted hydrogen-bond networks. Notably, the curved interface of the droplet displays an even higher population of under-coordinated motifs -- including 2D-1A, 1D-2A, and 1D-1A -- relative to the planar air-water slab (cf. Fig.~\ref{Figure1}c and Fig.~\ref{Figure1}d). Despite this quantitative enhancement, the qualitative evolution of hydrogen-bonding populations remains consistent across the two interfaces, indicating a common underlying interfacial structuring mechanism. Hydrogen-bond imbalances directly translate into local charge transfer. As discussed in Ref.~\cite{benamotz_science22}, configurations with excess donor bonds tend to become negatively charged, whereas those with excess acceptor bonds acquire a positive charge. Tables~\ref{tab:hb_family_meanCT} quantify this effect for droplet (left) and slab (right) geometries, while the underlying methodology is discussed in SI. It is worth stressing that a negative (positive) mean charge transfer corresponds to an electron depletion (accumulation), hence red (blue) entries represent positively (negatively) charged H$_2$O molecules. A clear trend emerges: hydrogen-bond donor-rich configurations (e.g., 1D-0A, 2D-0A, 2D-1A) exhibit electron accumulation (blue entries), while acceptor-rich ones (e.g., 0D-1A, 0D-2A, 1D-2A, 1D-3A) show electron depletion (red entries). Nearly symmetric configurations such as 2D-2A remain essentially neutral. This finding is in line with recent machine learning simulations of the charge transfer mechanism at the flat air-water interface \cite{Wang_JCP25}, while the close similarity between droplet and slab results indicates that this relationship is largely independent from the curvature.
\begin{table}[h]
\centering
\begin{minipage}{0.48\textwidth}
\centering
\begin{tabular}{ccc}
\toprule
$n_{\mathrm{don}}$ & $n_{\mathrm{acc}}$ & mean CT ($e^{-}$) \\
\midrule
0 & 1 & \cellcolor[rgb]{1.000,0.383,0.383} -0.052 \\
0 & 2 & \cellcolor[rgb]{1.000,0.041,0.041} -0.082 \\
1 & 0 & \cellcolor[rgb]{0.475,0.475,1.000}  0.045 \\
1 & 1 & -0.009 \\
1 & 2 & \cellcolor[rgb]{1.000,0.612,0.612} -0.033 \\
1 & 3 & \cellcolor[rgb]{1.000,0.652,0.652} -0.030 \\
2 & 0 & \cellcolor[rgb]{0.150,0.150,1.000}  0.072 \\
2 & 1 & \cellcolor[rgb]{0.747,0.747,1.000}  0.022 \\
2 & 2 & 0.000 \\
2 & 3 &  0.009 \\
\bottomrule
\end{tabular}
\end{minipage}
\hfill
\begin{minipage}{0.48\textwidth}
\centering
\begin{tabular}{ccc}
\toprule
$n_{\mathrm{don}}$ & $n_{\mathrm{acc}}$ & mean CT ($e^{-}$) \\
\midrule
0 & 1 & \cellcolor[rgb]{1.000,0.417,0.417} -0.050 \\
0 & 2 & \cellcolor[rgb]{1.000,0.000,0.000} -0.086 \\
1 & 0 & \cellcolor[rgb]{0.489,0.489,1.000}  0.043 \\
1 & 1 & -0.006 \\
1 & 2 & \cellcolor[rgb]{1.000,0.582,0.582} -0.036 \\
1 & 3 & \cellcolor[rgb]{1.000,0.709,0.709} -0.025 \\
2 & 0 & \cellcolor[rgb]{0.125,0.125,1.000}  0.074 \\
2 & 1 & \cellcolor[rgb]{0.711,0.711,1.000}  0.025 \\
2 & 2 & -0.001 \\
2 & 3 &  0.009 \\
\bottomrule
\end{tabular}
\end{minipage}
\caption{\scriptsize
Mean charge transfer (CT) associated with different hydrogen-bond donor-acceptor families for interfacial molecules in the droplet (left) and planar slab (right) geometry. Positive values (blue entries) indicate electron accumulation while negative values (red entries) indicate electron depletion. Negligible CT values are not colored.
}
\label{tab:hb_family_meanCT}
\end{table}

This way, the outermost interfacial layer exhibits a heterogeneous and electronically polarized environment, populated by locally charged molecular configurations. In particular, molecules exposing free OH groups toward the vapor phase tend to carry a positive excess charge, whereas configurations exposing lone pairs toward the air phase correspond to negatively charged H$_2$O molecules. The interface thus emerges as a fluctuating ensemble of oppositely charged environments arising from hydrogen-bond topology variations.

\subsection{Curvature Effects on the Interfacial Electric Field}

Although the two dominant charged configurations (2D-1A and 1D-2A) populate similar interfacial regions (Figs.~\ref{Figure1}c,d) -- thus ensuring overall charge neutrality and precluding any macroscopic capacitive charge separation -- the preferential orientation of water dipoles toward the liquid phase, combined with locally polarized and charged H$_2$O species, generate a finite interfacial electric field (IEF). 

\begin{figure}[h]
\centering
\includegraphics[width=\linewidth]{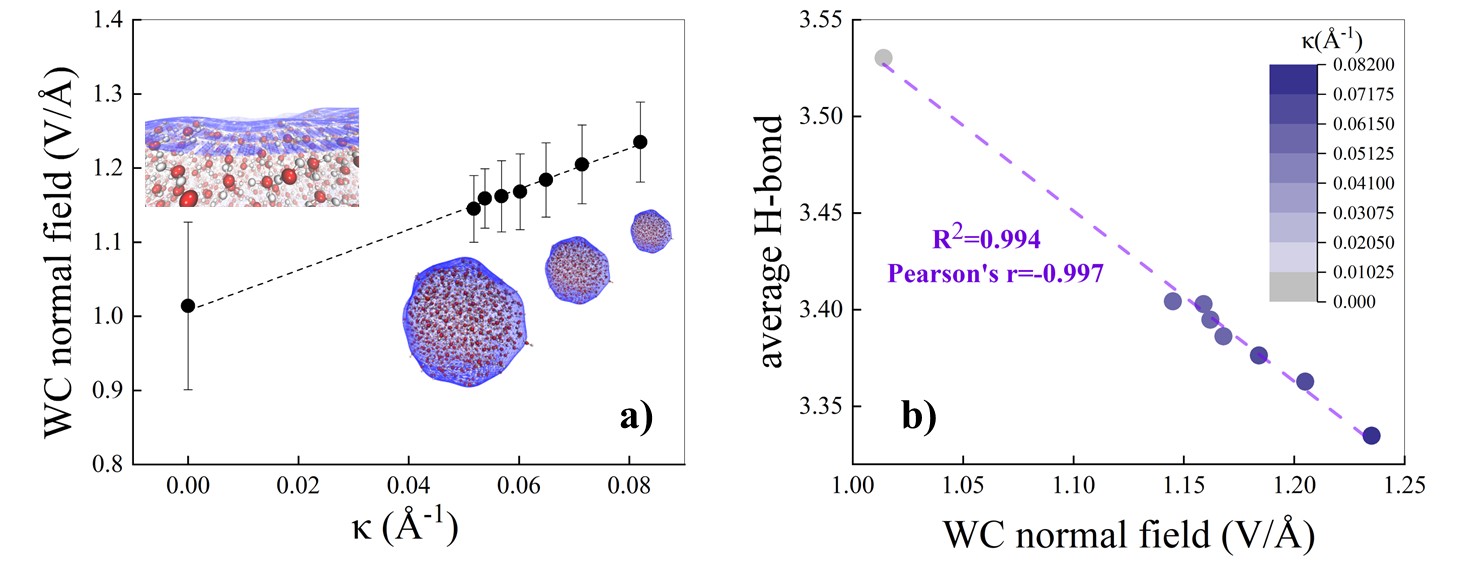}
\caption{\scriptsize (a) Average interfacial electric field projected along the Willard-Chandler surface normal as a function of the interface curvature $\kappa$ for droplets of different sizes and for the planar slab limit, whose structures are depicted as insets. (b) Correlation between the interfacial electric field and the average number of hydrogen bonds per interfacial molecule. The color scale indicates the corresponding curvature. Dashed lines are linear fit of the data points.
}
\label{Figure3}
\end{figure}
Fig.~\ref{Figure3}a reports the average electric field produced by the interfacial region and projected along the local WC surface normal. For all systems considered here, the field is positive, indicating an outward orientation from the liquid toward the vapor phase. Its magnitude ranges between $\sim$1.0 and 1.2 V/{\AA}, \emph{i.e.}, about 5-6 times larger than previous computational estimates\cite{Headgordon_NatCommun2022}. This discrepancy arises because the electric field, being a vector quantity, depends critically on how it is probed: earlier studies typically projected it along OH bond directions, which are generally less aligned with the local field than the WC surface normal adopted here.

Fig.~\ref{Figure3}a shows a remarkable linear dependence of the IEF on the curvature of the interface. As the curvature $\kappa$ increases, corresponding to smaller droplets, the magnitude of the field slightly increases. Conversely, as the droplet size increases and the curvature decreases, the field gradually approaches the value observed in the planar slab limit of zero-curvature, which represents the infinite-radius case. Although the trend with curvature is systematic, the overall variation of the IEF remains relatively small across the range of droplet sizes considered here. 
Recent experimental studies have reported significant differences in chemical reactivity in water microdroplets with diameters ranging from approximately 3 to 40~$\mu$m and have attributed this effect to variations in the IEF~\cite{XiaPNAS25,ZareJACS25}. Our results indicate that, at least from a structural and electrostatic perspective, curvature alone produces only marginal variations in the interfacial field strength in the experimentally probed curvature regimes. Extrapolation of the curvature dependence of the IEF to micrometer-sized droplets (3-40~$\mu$m) reveals variations on the order of $10^{-5}$ V/{\AA}, \emph{i.e.}, five orders of magnitude smaller than the field itself. This suggests that other factors than the electric field contribute more prominently to the experimentally observed reactivity differences.

To gain further insight into the microscopic origin of the IEF, Fig.~\ref{Figure3}b correlates the field intensity with the average number of hydrogen bonds per interfacial molecule. A clear anticorrelation is observed: as the hydrogen-bond network becomes more complete, the magnitude of the electric field decreases. In other words, droplets with lower curvature up to the slab limit exhibit a more complete hydrogen-bond structure and correspondingly weaker IEFs. This behavior highlights the intimate connection between hydrogen-bond topology and the electrostatic properties of the air-water interface. In highly curved droplets the hydrogen-bond network is more disrupted, leading to a larger population of under-coordinated molecules, larger polarization and charge transfer phenomena, and hence a stronger net electric field. As the interface becomes progressively flatter, the hydrogen-bond network becomes more structured and the interfacial field correspondingly weakens. Nevertheless, even in the slab limit the hydrogen-bond coordination remains lower than in the bulk phase (see, \emph{e.g.}, Table S2 of the SI), reflecting the intrinsically incomplete hydrogen-bond network of the liquid-vapor interface highlighted in Fig.~\ref{Figure1}.

\subsection{pH Effects on the Interfacial Electric Field}

Several experimental methodologies exist to produce water microdroplets, some of them giving rise to globally neutral droplets while others, most notably electrospray ionization (ESI), generate highly charged droplets through the application of strong voltages \cite{Fenn1989, Banerjee_AnalChem2012}. Microdroplets can also be produced via gas nebulization \cite{Gao2019}, ultrasonic humidification \cite{Mishra2021,Zare_JACS2022}, vapor condensation \cite{Lee2019}, droplet deposition on surfaces \cite{Wei2021}, or levitation techniques \cite{Bain2016}. These methods rely on distinct mechanisms, all involving significant energy input (electric fields, sound, or heat) that perturbs water’s physico-chemical state. ESI can generate droplets with excess \ce{H3O+} or \ce{OH-} without counterions \cite{Banerjee_AnalChem2012}. Even without external fields, nebulization and sonication can yield charged droplets via contact electrification or ion partitioning \cite{LaCour_JACS25}, thereby shifting the autoprotolysis equilibrium and promoting excess ionic species. Given the catalytic activity of such charged nanodroplets, disentangling the roles of excess charge and IEFs becomes essential.
\begin{figure}[t!]
\centering
\includegraphics[width=0.9\linewidth]{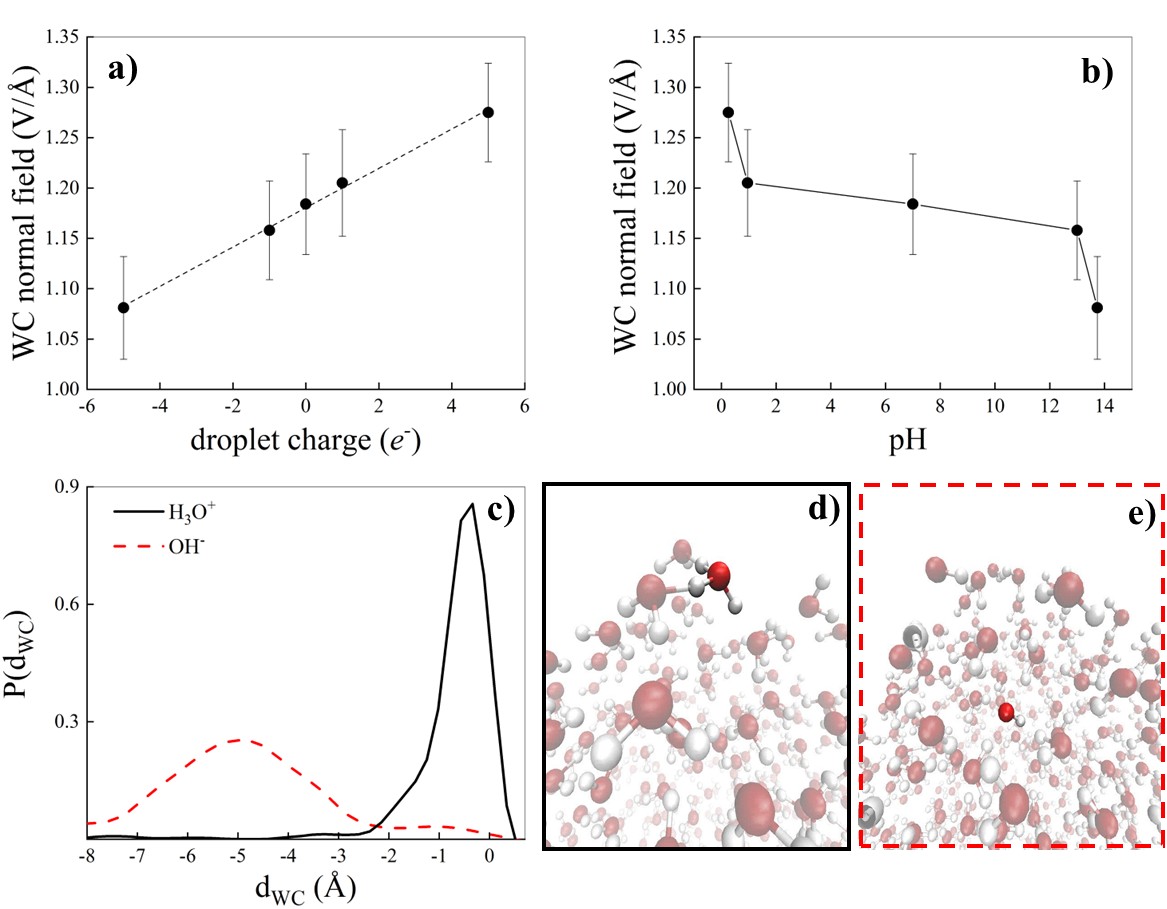}
\caption{\scriptsize 
Average interfacial electric field projected along the Willard-Chandler surface normal as a function of the droplet charge (a) and resulting pH (b). In panel (a), the dotted line is a linear fit of the data points. In panel (c), the distributions of the distances of an excess hydronium (solid black line) and hydroxide (dashed red line) species with respect to the Willard-Chandler instantaneous surface in nanodroplets containing 511 H$_2$O molecules and one excess ion are displayed. Panels (d) and (e) depict typical molecular arrangement of the hydronium and hydroxide water counterions, respectively, extracted from our deep potential molecular dynamics simulations.}
\label{Figure4}
\end{figure}

To quantify the impact of water counterions on the IEF, we simulated several nanodroplets with excess \ce{H3O+} or \ce{OH-}. As detailed in the Methods section, we simulated charge states corresponding to $-5e^-$, $-e^-$, $+e^-$, and $+5e^-$, hence probing regimes of $13\%$ and $66\%$ of the Rayleigh limit for Coulomb fission. These charge states approximately correspond to net pH values of 13.74, 13.0, 0.96, and 0.26, respectively. Figure. \ref{Figure4}a shows that the IEF linearly increases with the total excess charge, suggesting that \ce{H3O+} and \ce{OH-} ions act effectively as point charges in shaping the interfacial field. Due to its logarithmic definition, a different trend emerges when the electric field is plotted as a function of the nominal pH (Fig. \ref{Figure4}b). It turns out that the most significant variations of the electric field occur at extremely acidic or basic pH values. Interestingly, whereas at acidic pH values the projection of the IEF on the WC surface normal increases, in basic conditions it measurably decreases with respect to the neutral case. This observation reconciles with several factors governing the nature of the hydronium and the hydroxide ions. Firstly, a positively charged ion (\ce{H3O+}) localized at the interface becomes an additional field source. On the contrary, an anion (\ce{OH-}) acts a sink of the electric field lines. Secondly, the profoundly different chemical nature of the water counterions leads to a sizably different behavior at the interface. In fact, whereas \ce{H3O+} is a very good hydrogen-bond donor (\emph{i.e.}, up to three) but a poor acceptor, the \ce{OH-} moiety can donate up to only one hydrogen bond but is capable of accepting even five hydrogen bonds and lying in over-coordinated states\cite{Chen2018}. Those features confer different affinities of those ions toward the interface, as widely debated in literature \cite{Jungwirth2006,McCaffreyPNAS2017,SekiJACS2023}. In particular, hydronium tends to localize in the topmost layer of the droplet, whereas the hydroxide exhibits the tendency to reside slightly deeper in layers between the interface and the bulk, as shown by the distributions of Fig. \ref{Figure4}c and in the snapshots of Figs. \ref{Figure4}d,e. In line with other computational evidences stemming from reactive classical force fields\cite{Headgordon_NatCommun2022}, the emerging picture suggests only a modest variation of the IEF achieved at extreme pH values. This neatly contrasts with the idea that the electric field itself might be the carrier for the observed enhanced catalysis as a function of the applied voltage in, \emph{e.g.}, electrosprayed charged droplets\cite{Banerjee_AnalChem2012}.   

\subsection{Spatial Extension of the Interfacial Electric Field}

\begin{figure}[t!]
\centering
\includegraphics[width=\linewidth]{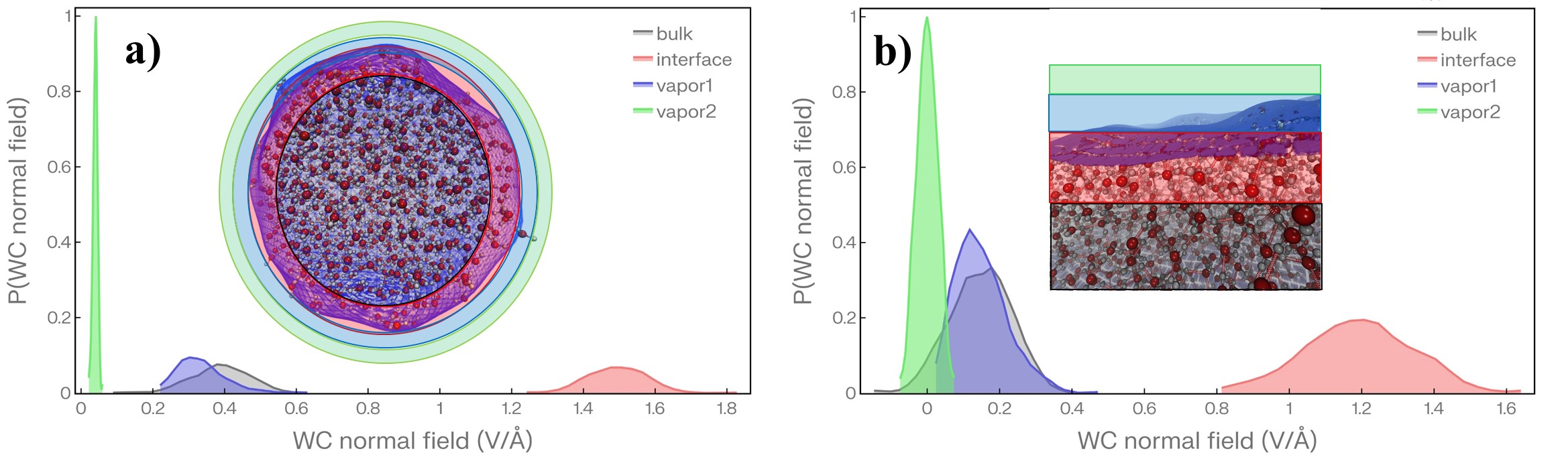}
\caption{\scriptsize Histograms of the electric field projected along the Willard-Chandler surface normal in different spatial regions. Panel (a) shows the nanodroplet geometry, while panel (b) shows the planar slab interface, as also visualized in the respective insets. Histograms correspond to bulk (gray), interface (red), near-vapor (blue), and far-vapor (green) regions. See text for characterization of these regions.
}
\label{Figure5}
\end{figure}

Fig.~\ref{Figure5} reports the distribution of the electric field projected along the WC surface normal across distinct spatial regions relative to the interface, enabling a quantitative assessment of its evolution with distance. Four regions are identified: bulk, interfacial, near-vapor (vapor1), and far-vapor (vapor2). The bulk region extends from the liquid interior up to 2.6 and 3.4~{\AA} below the WC instantaneous interface in the droplet and slab systems, respectively. The interfacial region spans from this threshold to 0.8~{\AA} (0.2~{\AA}) above the WC surface in the droplet (slab), thereby including all molecules residing at the interface. This difference arises from the distinct positioning of the WC surface in curved and planar geometries, as also reflected in the density profiles reported in Fig.~S1 of the SI. The vapor region is further subdivided into two adjacent layers: vapor1, with a thickness of 2.0~{\AA} (2.6~{\AA}) in the droplet (slab), and vapor2, 1.2~{\AA} thick in both systems, both located above the interface, as illustrated in the insets of Fig.~\ref{Figure5}.

In both droplet (Fig.~\ref{Figure5}a) and slab (Fig.~\ref{Figure5}b) geometries, the largest electric field values are found within the interfacial region. The corresponding distributions peak at approximately $1.2$--$1.4$ V/{\AA}, strengths slightly higher than those reported in Figs.~\ref{Figure3}-\ref{Figure4}. This difference arises from the refined definition of the interface adopted here, which excludes lower-field contributions originating from vapor-phase molecules and from regions of vacuum located $\gtrsim 1$ {\AA} above the WC surface, which fall in the vapor1 and vapor2 regions. As anticipated, the combined effects of preferential dipolar orientations, asymmetric hydrogen-bonding environments, and enhanced charge fluctuations associated with under-coordinated interfacial molecules give rise to remarkably strong local electrostatic fields within the first molecular layers of the interface.

Moving away from the interface toward the vapor phase, the magnitude of the electric field very rapidly decreases. In the vapor1 region the typical field values drop to smaller values of approximately $\sim$ 0.3 V/{\AA}, while in the vapor2 region the field vanishes within the statistical uncertainty. The non-zero average field observed in the bulk originates from the artificial -- yet unavoidable -- sharp cutoff used to separate bulk and interfacial regions, which inevitably truncates part of the hydrogen-bond network at the boundary. These results demonstrate that the IEF is localized and decays within only a few {\AA} from the interface. Beyond the first molecular layers, the electrostatic environment becomes comparable to that of the vacuum phase.

\section{Conclusions}

In this work, we have provided a fully quantum-mechanical characterization of the electric field at air-water interfaces across planar geometries, nanodroplets of varying curvature, and neutral and charged conditions. By combining neural-network-based deep-learning molecular dynamics with \emph{ab initio} electronic-structure re-sampling, we show that the interfacial electric field (IEF) is a robust feature of aqueous interfaces, consistently oriented outward and governed by the local hydrogen-bond network.

A central result is that both curvature and pH exert only a minor influence on the IEF. Although a systematic linear trend with curvature is observed, its magnitude remains limited, and extrapolation to micrometer-sized droplets reveals variations that are orders of magnitude smaller than the field itself. Similarly, while excess charge modulates the field in a predictable manner, its impact becomes significant only under extreme pH conditions. These findings indicate that the experimentally reported reactivity enhancements in microdroplets due to either curvature \cite{XiaPNAS25,ZareJACS25} or charge \cite{Banerjee_AnalChem2012} cannot be directly attributed to variations in the interfacial electrostatics, but must instead arise from alternative molecular mechanisms. Importantly, the electric field is highly localized, decaying within only a few {\AA} from the interface. Within this narrow region, it arises from interfacial molecular organization, including dipolar orientation, hydrogen-bond disruption, and charge-transfer fluctuations, while rapidly vanishing beyond the first molecular layers.

These results provide a revised perspective on interfacial electrostatics. The IEF emerges directly from the local hydrogen-bond topology, molecular orientations, and associated polarization and charge-transfer fluctuations that characterize the first interfacial layers. Its strong spatial localization, with a decay over only a few {\AA}, prevents the establishment of long-range electrostatic effects. 
Taken together, these findings indicate that the electric field is not an independent driving force for interfacial reactivity, but rather a direct, local, manifestation of the underlying molecular and electronic structure. Within this framework, the catalytic properties of aqueous interfaces are more naturally attributed to local chemical effects -- including changes in solvation structure and intermolecular charge transfer within the first interfacial layers -- thereby strongly challenging the common assumption that IEFs are primary driving forces of ``on-water'' catalysis.



\begin{acknowledgement}
G.~A. and G.~C. acknowledge support by ICSC - Centro Nazionale di Ricerca in High Performance Computing, Big Data and Quantum Computing, funded by European Union - NextGenerationEU - PNRR, Missione 4 Componente 2 Investimento 1.4. G.~C. acknowledges the European Union - NextGeneration EU  from the Italian Ministry of Environment and Energy Security POR H2 AdP MMES/ENEA with involvement of CNR and RSE, PNRR - Mission 2, Component 2, Investment 3.5 ``Ricerca e sviluppo sull’idrogeno'', CUP: B93C22000630006. G.~C. acknowledges the  European Union (NextGeneration EU), through the MUR-PNRR project SAMOTHRACE (ECS00000022). G.~C. is thankful to CINECA for awards under the ISCRA initiative, for the availability of high-performance computing resources and support. 
\end{acknowledgement}

\begin{suppinfo}


\end{suppinfo}

\bibliography{achemso-demo.bib}

\clearpage


\end{document}